# Interaction of Gold Nanoparticles in Barium Titanate Thin Films

*Yaodong Yang*, Jianjun Yao, Yu U. Wang, Jiefang Li, Jaydip Das, and Dwight Viehland*

Department of Materials Science and Engineering, Virginia Tech, Blacksburg, Virginia 24061, USA

* E-mail: yaodongy@vt.edu

ABSTRACT

A novel approach to control the grain size of oxide thin film materials has been investigated. Perovskite $BaTiO_3$ shows interesting grain structures when deposited on gold pre-deposited, (111) oriented, single crystal $SrTiO_3$ substrates. Solid oxide films grow epitaxially on patterned seed layers, and show variations in grain size relative to the films deposited on $SrTiO_3$ directly.

Piezoelectric oxide materials have drawn recent research interest due to their potential in many applications ranging from sensors to radio-frequency devices.[1-3] The piezoelectric properties depend on the polarization distribution of individual domains, which is affected by grain size and orientation.[4-6] Growth of piezoelectric thin films on particular substrates can enhance the utility of these materials for practical purposes due to the possibility for size and cost reduction, better compatibility, and improved device performance.[1] Recent developments in thin film technologies offer the opportunity to control the grain size and distribution precisely, and to perform a detailed study of the dependence of the piezoelectric response on grain structures. For practical on-chip applications, such films can be useful either as a continuous film or with well defined patterns. Patterns also can be made by, for example, focused ion beam which involves a tedious and expensive process.

Consider perovskite $BaTiO_3$ (BTO) thin films, for example, that exhibit good piezoelectric responses.[7-9] Epitaxial BTO films can be deposited by pulsed laser deposition (PLD) at a temperature of about 700°C or higher. To make BTO nano-structures, one has to pattern the film layer post-deposition: since the deposition temperature is too high for a polymer photo-resist as normally used in lithography techniques.

Here, we report a simple technique to pattern the BTO films directly during deposition on gold (Au) nanoparticle pre-deposited $SrTiO_3$ (STO) substrates. Here, the pre-deposited Au served as a seed layer that affects the growth mechanism of the BTO layer.[10] The BTO grain size and shape are found to depend on the Au nanoparticles. Note that, in addition to work as a patterned template, Au might be used for other purposes. For example, Au being a well known electrode material, can serve as embedded electrodes to apply electric voltage across the piezoelectric materials. Further, the Au-BTO can be considered as a metal-ceramic composite and it might be possible to study the interaction between metal and ceramic that can help us bring these advantages from each material to achieve a better performance.[11]

RESULTS AND DISCUSSION

Figure 1 illustrates the SEM results. Parts (a) and (b) show images for the patterned BTO films with Au pre-deposited seed layer. Parts (c) through (e) show images for BTO films deposited directly on the STO substrate with an attached TEM grid. From Fig. 1(a), we can clearly see that the BTO films have a hexagonal shaped pattern. These patterns look exactly similar to the patterns in the TEM grid through which the Au was sputtered. This indicates that the gird with hexagonal shaped windows first helps to pattern the Au layer, and then the Au layer works as a positive template to pass this pattern to the BTO layer by controlling the grain distribution.

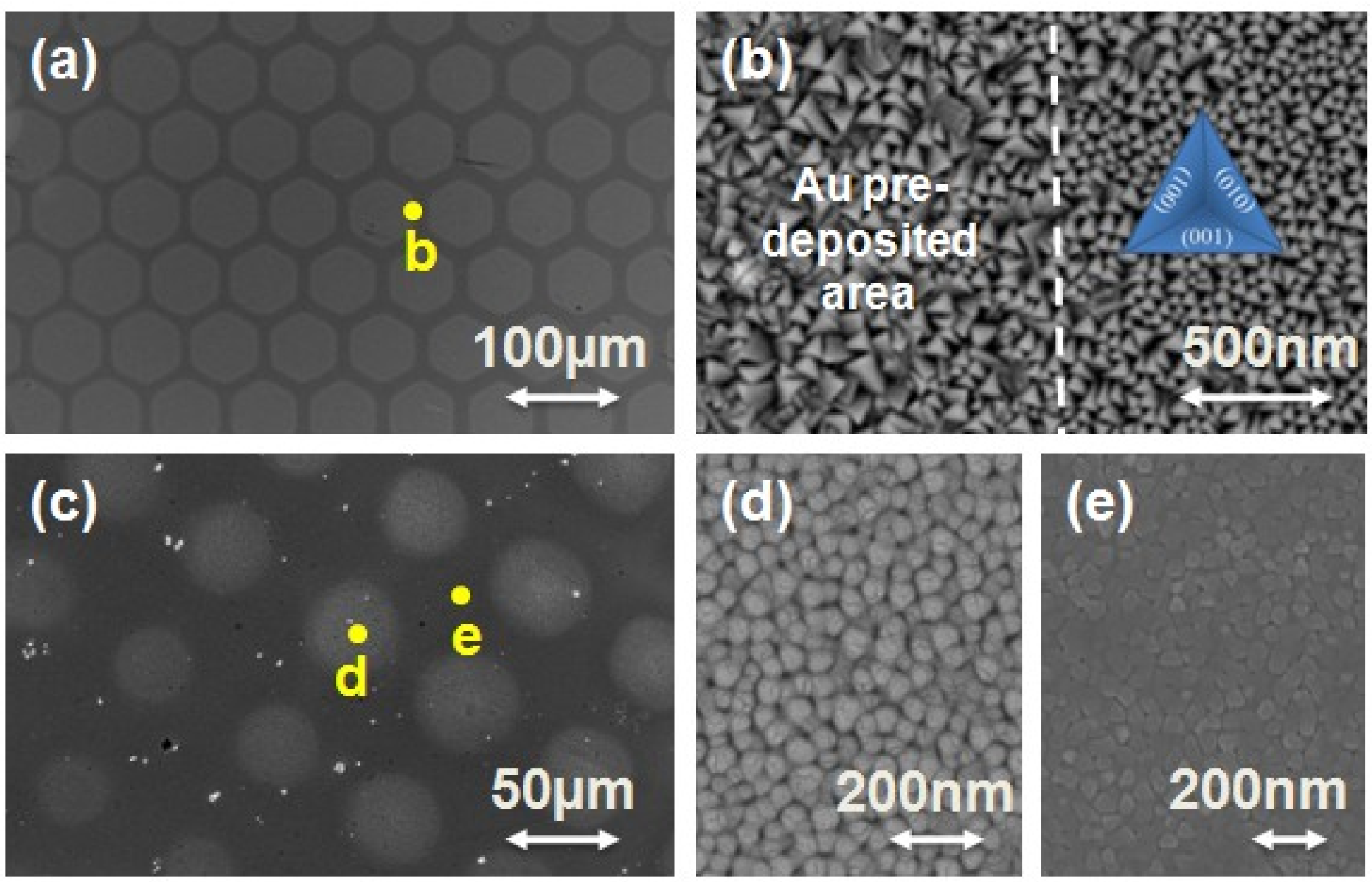

Figure 1. SEM images of (a) patterned Au pre-deposited BTO; (b) higher magnification image of the boundary area as marked in (a), where the left side is Au pre-deposited and patterned area; (c) directly patterned BTO thin film on STO substrate; and (d, e) higher magnification images of grid uncovered and covered areas as marked in (c).

To study the boundary in more detail, we obtained SEM images of the positive/negative patterned area (marked by a dot "b" in Fig. 1(a)), as shown in Fig. 1(b). A dashed line is used to illustrate this boundary. A notable difference between the Au pre-deposited areas can be found: the grain size in the positive patterned (i.e., Au pre-deposited) area was approximately 100 nm, whereas that in the negative patterned (exposed) area was approximately 50 nm. Furthermore, one can see that both areas have pyramidal grain morphologies that are indicative of (111) oriented BTO grains. A schematic diagram in Fig. 1(b) illustrates the BTO unit cell projected onto the (111) plane, representing the pyramidal shape.

Next, we obtained images of the BTO morphology when deposited directly on STO without an Au seed layer as a control sample. In this case, we put the copper grid in the chamber and deposited BTO directly onto the STO substrate through it. From Fig. 1(c), it can clearly be seen that the BTO film morphology appears circular instead of hexagonal, and that there was no clear boundary that developed. We also obtained images under higher magnification of both uncovered and covered areas, as shown in Figs. 1(d) and (e). Both areas had BTO films, and the difference was that there was more BTO in the uncovered window areas. Note that the grain size and shape are nearly the same for both areas. We also noticed a slightly difference between the right hand side of Fig.1(b) and Fig.1(d): the triangular grains in Fig.1(b) are sharper than that in Fig.1(d), even though the grains are in the same size. This difference may have resulted from the presence of some Au deposition beneath the masked areas at the boundary. Small Au articles could affect the BTO grain growth after dewetting.

Based on the above results, it looks like the Au seed layer seemed to play an important role in affecting the grain growth in the BTO layer. In order to understand the effect of the Au seed layer on the BTO growth mechanism better, we prepared a series of control samples. Figure 2a shows BTO films deposited directly on STO substrate (without Au seed layer) under the same conditions, as mentioned above. Part (b) shows a SEM image taken from an as-prepared Au layer deposited by Argon sputtering at room temperature for 20seconds. This layer looks more or less continuous with several randomly distributed holes that are tens of nanometers in size. Part (c) is an image showing "hundreds of nanometer large" Au islands obtained after annealing at 750°C for 40mins in a vacuum chamber at an oxygen pressure of 100 mTorr. Since this is a control experiment, the annealing process was performed in the same manner as the PLD deposition procedure, but without turning on the laser. From the image we can see numerous island-like formations that are somewhat regularly spaced: some of which are circular in shape and others which are trilateral in shape. This island-like formation in Au on the STO substrate can be attributed to the difference in the surface energies and diffusion kinetics between the metal (Au) and oxide (STO) and the growth and coalescence of Au particle at such annealing temperatures.[12] Part (d) shows the PLD-deposited BTO film morphology on the Au seeded STO substrate, which had larger grain sizes than the BTO film grown directly on STO (see Fig.2a). The findings for this control experiment in Figure 2 clearly demonstrate that the Au seed islands on the STO substrate helps to grow the BTO grains faster, resulting in larger grains than the BTO grown directly on the STO substrate.

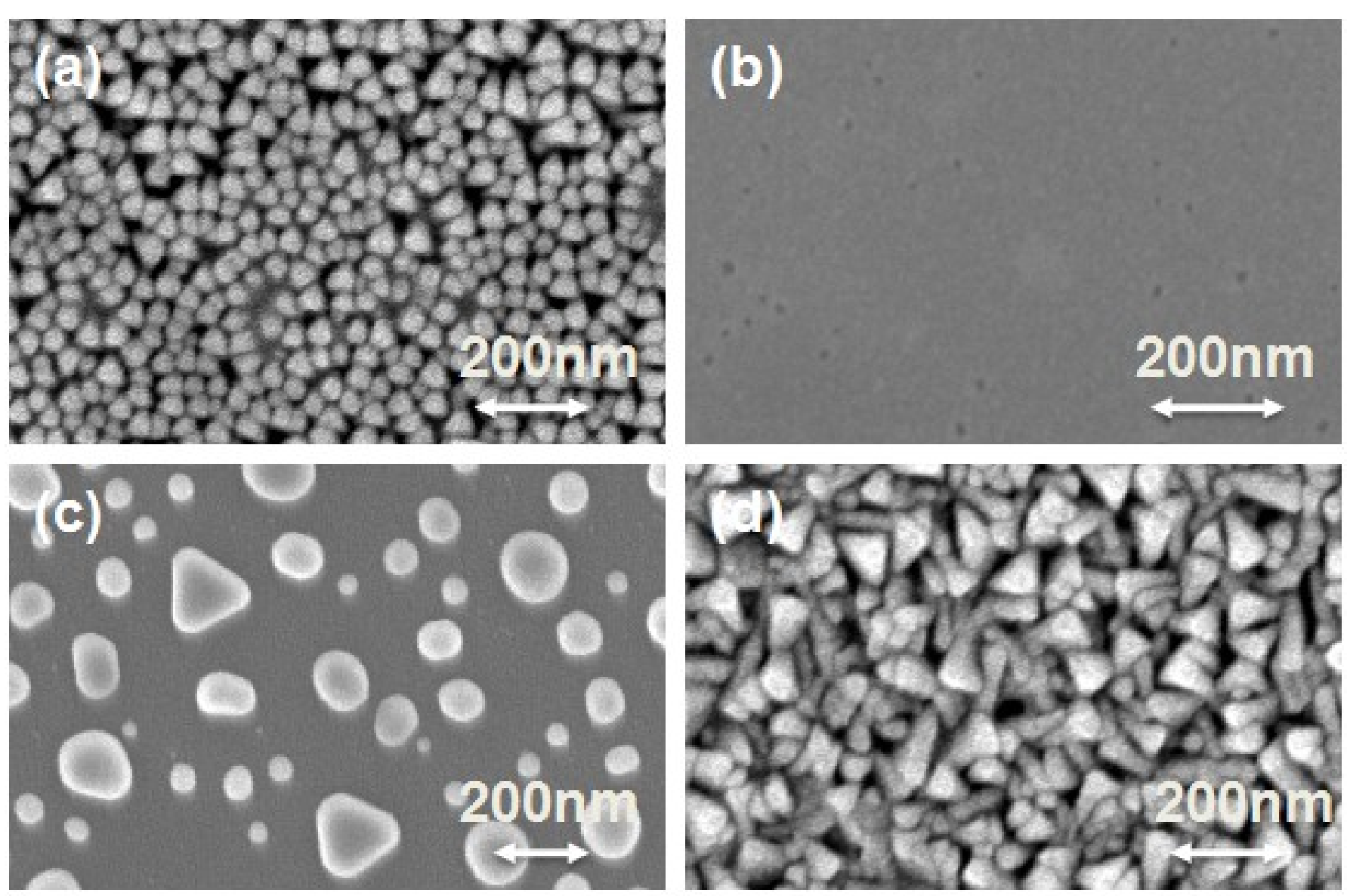

Figure 2. SEM images of (a) BTO thin film deposited directly onto a (111) oriented STO substrate by PLD; (b) Au film deposited by sputtering for 20 seconds at room temperature; (c) same sample shown

in (b) after annealing at 750$^{o}$C for 40minutes; and (d) BTO thin film with Au pre-deposited layer showing larger grain sizes than when grown directly on STO (see Fig.2a).

We then used FIB to lift out a small cross sectional piece as a TEM sample. HRTEM images in Fig. 3 provide more detailed information concerning the morphology and grain size of the BTO structure that was grown on the Au pre-deposited substrate. Figure 3a shows the Au nanoparticles and the BTO grains. One can see that the Au formed nanoparticles of two different sizes: the dominant one was approximately 40-50 nm in size and was distributed more or less in a regular fashion, and the minor one was approximately 15 nm in size and was occasionally observed in the space between major particles. This indicates that the hexagonal shaped Au-layer internally breaks up to into near regularly spaced Au islands, possibly on heating the substrate.[10, 13] From the image, we can also see that the BTO nucleated and grew from the dominant Au nanoparticles, where each metal nanoparticle supports one BTO columnar grain: the regularity of the Au nanoparticle spacing seemed to be passed onto the BTO columns. Two lines are drawn in the image to serve as a guide for the eyes to follow the BTO columnar grains grown on Au nanoparticles. The inset scanning TEM (STEM) image shows that there are notable gaps between the BTO grains.

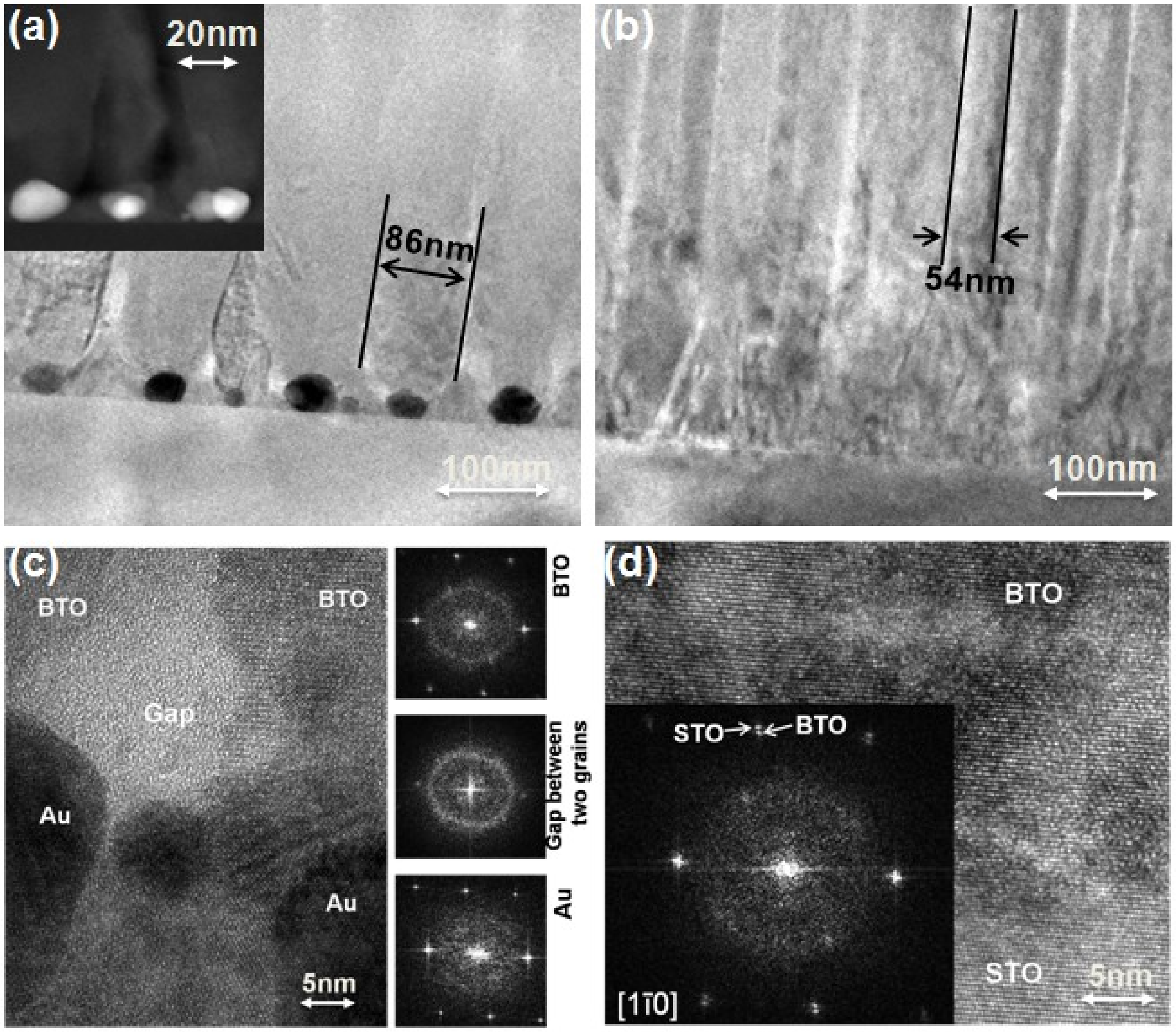

Figure 3. TEM images (a, b) and HRTEM images (c, d) of BTO thin film to reveal the interphase interfacial areas with (a, c) and without (b, d) Au-seed layer, respectively. Inset in (a) is a STEM image taken from the same area as that of the TEM image; inserts in (c) are power spectra taken from Au, gap and BTO areas, respectively, where the corresponding areas from which these spectra were taken are marked in the TEM image, and the zone axis is same. Inset in (d) is a power spectrum taken from the entire area shown in the image.

Figure 3(b) shows a HRTEM of the BTO film that was deposited directly on a STO substrate without seed layer. Again, one can see columnar grain growth, as marked by two lines in the image. There were two noticeable differences between the images in Figs. 3(a) and 3(b). First, the BTO grain size (the lateral distance between the lines) was about 90 nm for the Au pre-deposited film (as determined by the distance between Au nanoparticles), whereas for the areas without any Au beneath, it was about 50nm. This is in agreement with the SEM images shown in Fig. 1. Second, the BTO-STO interfacial area for the film grown with Au was more ordered than that for the other film. The BTO grain boundaries are quite visible just above the Au nanoparticle in Fig. 3(a); while in Fig. 3(b), the BTO columnar grains could be clearly seen at only about 100 nm above the substrate surface. This demonstrates that the Au nanoparticles significantly reduce the interfacial stress in the BTO and thus, help to relax the lattice distortion in BTO much faster. Elemental analysis by EDS (attached to the Titan HRTEM system) confirmed that the areas discussed above were indeed BTO and Au (Fig.S1).

Figure 3(c) shows a higher magnification image of the grain boundary of Fig. 3(a). It can be seen that the BTO grains grown on the Au nanoparticles are single crystalline, which is marked by "BTO" in the image, as determined by power spectra to have the same orientation as the STO substrate. Between two oriented BTO grains, there was a small transition area that corresponded to an amorphous phase, which was marked as "gap" in the image. The Au nanoparticles were crystalline as evidenced by power spectra. Clearly, the BTO columnar grains nucleated and grew from the crystalline gold nanoparticles in an orderly manner. However, a close look at the interphase interface of the BTO film shows that BTO is reasonably epitaxial with the STO substrate, as shown in Fig. 3(d). The inset shows spot splitting in a power spectrum, which proves epitaxial growth from the STO.

The SEM and TEM images shown in Figs. 2 and 3, respectively, indicate that Au-buffer layers serve as nucleation sites for the formation of BTO grains. As evident from the images, without the buffer layer, the BTO interfacial region is notably stressed. This compressive stress presumably comes from the 2.5% lattice mismatch between the BTO film and the STO substrate. The BTO nuclei crystallize on the STO substrate, and subsequently merge together during grain growth. To minimize interfacial energy between grains and the substrate, grains grow into a columnar morphology with increasing thickness.

For BTO films deposited directly on STO substrates, elastic stress is relaxed with columnar grain growth, producing quite regular interfaces between BTO grains at film thickness above about 100 nm.

For the case of Au pre-deposited BTO, the growth process is more complicated. The lattice parameter for cubic Au is about a=4.07Å, while for STO is about a=3.90Å. BTO is tetragonal, with lattice parameters of ($a_t$, $c_t$)=(3.99 Å, 4.01 Å). Interestingly, the lattice parameters of BTO lie between those of Au and STO, which is a little closer to that of Au; thus using STO-Au as a substrate will match the BTO lattice parameter better than only STO or Au substrates. Figure 3(c) shows that Au is epitaxial with the STO substrate, which would reduce the lattice parameter of Au nanoparticles by epitaxial stress to become even closer to that of BTO. This may be a reason for preferred BTO nucleation on Au nanoparticles, which helps to relax the stress imposed on BTO from the STO substrate. Accordingly, this would provide a way for better epitaxy, and the lattice mismatch is accommodated by the amorphous areas between BTO grains in the vicinity of Au nanoparticles. As a result, BTO grain boundaries are well formed at a much smaller distance from the substrate surface. From Figs. 3(a) and (c), it can clearly be seen that ordered BTO grains grow directly from the Au nanoparticles, and that the grain size is determined by the distance between nucleation sites (i.e., Au nanoparticles). In the regions between Au nanoparticles with exposed STO substrate surface, a triangular shaped small gap region of amorphous BTO existed that reduced the interfacial stress between gold, BTO, and STO. As the array of BTO grains on Au nanoparticles began to grow longer, this gap area was closed, leaving only a residual triangular amorphous region at the interfaces where all three phases were in close proximity, as shown in Fig. 3 (a).

PFM was then used to investigate the ferroelectric domain structures in the Au pre-deposited BTO thin films in order to study the effect of grain size on the physical properties. For this purpose, the Au-buffer layer and BTO film was deposited on conducting Nb (0.5 wt %) doped STO substrates, which serve as the bottom electrode for the PFM measurements. The topographic AFM image is shown in Fig.

4(a) and the corresponding piezoresponse amplitude and phase images are listed in (b) and (c). As shown in Figs. 4(a), (b) and (c), the surface morphology and piezoresponse image was composed of two regions, indicated by white lines. The left side was smoother than the right side (Au pre-deposited), and had a lower degree of roughness and smaller grain size. The difference in height between these two regions was approximately 60 nm, indicating that BTO grains grow faster on Au nanoparticles, which also supports our observations from SEM and TEM images. The PR amplitude data is shown in bigger and darker triangles to indicate stronger piezoresponses, and PR phase signal is shown in darker colors to better reveal areas with stronger piezoresponses, as shown in Fig. 4(b) and (c). This image shows that the PR signal (both phase and amplitude) on the left side does not have significant contrast, indicating that the film is not strongly piezoelectric. However, on the right side of this image, the triangular BTO grains show pronounced contrast, especially at the top of the triangles. The size and density of dark regions on the right side of the image were both greater than on the left. The inset of Fig. 4(b) and (c) enlarges several grains. In this inset, we can see that the piezoresponse also had a uniform triangular shape, indicating that the PR in one triangular grain is from a single domain. These PFM images show that the domain distribution of films grown on Au pre-deposited areas is much more like a single domain single crystal state than the areas grown directly on STO.

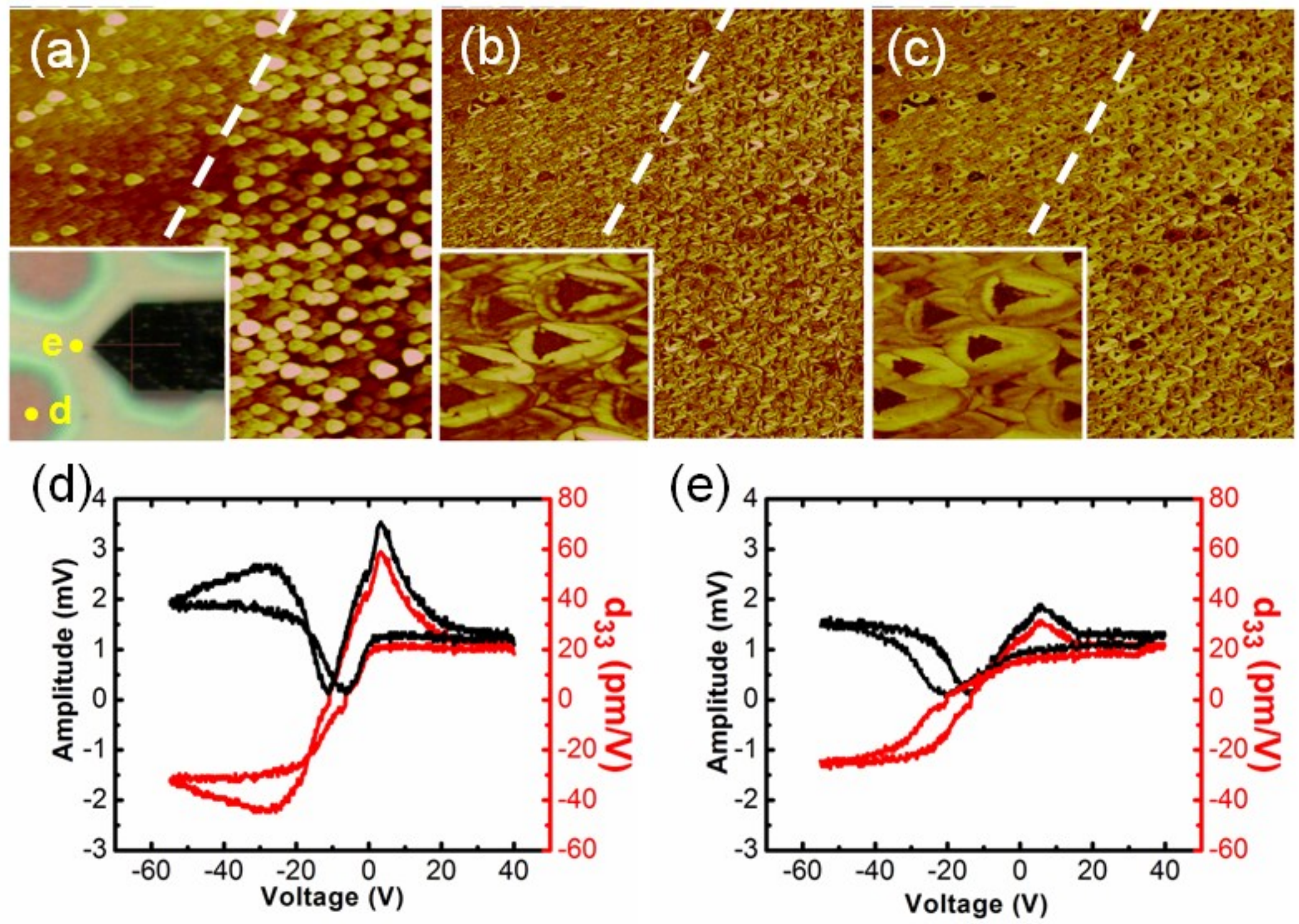

Figure 4. AFM and PFM study of transition zone: (a) AFM topography image of an area of 20 $\mu$m $\times$ 20 $\mu$m, where the insert is an optical view of the scanned area; (b and c) the corresponding piezoresponse amplitude image and piezoresponse phase image, where the insets are the enlarged image of the area with three triangular grains grown on Au pre-deposited STO; and (d) and (e) are the local piezoelectric hysteresis loops measured in and out Au pre-deposited areas (marked by “d” and ”e” as in Fig.4a inset), respectively.

In order to get a better insight into the local piezoelectric properties of the two different regions, piezoelectric hysteresis loops were measured under the same AC bias conditions (3V, 20 kHz). The measurement used a conductive cobalt coated cantilever with a spring constant of 0.0678 N/m. The results can be seen in Fig.4 (d) and (e). A voltage shift of the loops towards a negative bias was found: this kind of shift has been attributed to the formation of space charges at the film/electrode interface.[14] The coercive voltage of the Au pre-deposited area (5V) was a little smaller than the other region (8V). An asymmetric shape for the loops was observed, which may have resulted from higher polarization under positive bias.[15] Meanwhile, the average amplitude of the Au pre-deposited area and corresponding piezoelectric coefficient was also greater than that of the other regions. The magnitude of $d_{33}$ in the Au pre-deposited area reached up to 60 pm/V; however out of the Au pre-deposited areas, the peak value was about 30 pm/V. This indicates that the Au pre-deposited area with larger grain sizes at the nanoscale have better piezoelectric properties. We believe that the increase of grain boundary area on the region without Au might degrade the piezoelectric properties of the BTO film.

## CONCLUSIONS

In summary, Au nanoparticles have been shown to be capable of controlling the grain size of BTO films, without change in the epitaxial growth conditions of the thin films. These findings should allow for enabling control of piezoelectric properties *via* domain engineering. Since Au nanoparticles show ferromagnetism for particle sizes smaller than 10 nm,[16-18] these findings also establish a relationship between BTO and Au, which may allow for novel magnetoelectric interactions.

## METHODS

In this investigation, we used a transmission electron microscopy (TEM) grid (300mesh copper grid from SPI company) with hexagonal windows as a template to sputter Au thin layers onto (111) oriented STO single crystal substrates at room temperature for 20 seconds. The current used to sputter Argon was 45mA. The TEM grid was subsequently removed. A pulsed laser deposition (PLD) system with a KrF laser of wavelength 248nm (Lambda Physik 305i) was used to deposit a BTO film on the patterned Au-pre-deposited STO substrate. Deposition was done in a vacuum chamber at an oxygen pressure of 100 mTorr. The distance between the substrate and target was 8 cm and the substrate was heated to 750°C. A laser spot of 3 $mm^2$ size and 1.2 J /$cm^2$ energy density was rastered at a frequency of 10 Hz on a stoichiometric target surface for 40 min. After deposition, the films were cooled under a 760 Torr oxygen pressure to room temperature. For comparison purposes, we attached the same TEM grid to the STO substrate and directly deposited BTO films on the STO substrate at the same temperature.

Scanning electron microscopy (SEM) images and energy dispersive spectrums (EDS) were obtained using a LEO (Zeiss) 1550 high-performance Schottky field-emission SEM. A FEI Helios 600 NanoLab FIB SEM was used to prepare and lift-out TEM samples. A FEI Titan 300 high-resolution TEM (HRTEM) was used to obtain lattice images and high-resolution EDS. Atomic force (AFM) and piezoelectric force (PFM) microscopy images were obtained by a Veeco Dimension 3100 Nanoman AFM.

ACNOWLEDGEMENTS

Support for this work was provided by the Division of Materials Research of the National Science Foundation. Authors also give thanks for NCFL in Virginia Tech for all the SEM, FIB and TEM support.

*Supporting Information Available*: Elemental analysis by EDS (attached to the Titan HRTEM system) is available free of charge *via* the Internet at http://pubs.acs.org.